\documentclass[epjCONF, columns]{svjour}
\usepackage{graphics}
\usepackage{slashed}				%
\usepackage{amsmath}				%
\usepackage{multirow}				%
\usepackage{subfigure}				%
\usepackage{hyperref}				%
\usepackage[font=small,labelfont=bf]{caption}	%
\usepackage[varg]{txfonts} % Times fonts
\usepackage[latin1]{inputenc}
\session-title{Hadron Collier Physics Symposium 2011, Paris, France}
\begin{document}
\title{E$_6$SSM vs MSSM gluino phenomenology}
\author{\underline{Patrik Svantesson}\thanks{Presenter of the project at HCP-2011}\thanks{\email{\href{mailto:P.Svantesson@soton.ac.uk}{P.Svantesson@soton.ac.uk}}}\and Alexander Belyaev\and Jonathan P.~Hall \and Stephen F.~King}
\institute{University of Southampton}
\abstract{
The E$_6$SSM is a promising model based on the group $E_6$, assumed to be broken at the GUT scale, leading to the group $SU(3)\times SU(2)\times U(1)\times U(1)'$ at the TeV scale. It gives a solution to the MSSM $\mu$-problem without introducing massless axions, gauge anomalies or cosmological domain walls. The model contains three families of complete 27s of $E_6$, giving a richer phenomenology than the MSSM. The E$_6$SSM generically predicts gluino cascade decay chains which are about 2 steps longer than the MSSM's  due to the presence of several light neutralino states. This implies less missing (and more visible) transverse momentum in collider experiments and kinematical distributions such as $M_{\mbox{\tiny eff}}$ are different. Scans of parameter space and MC analysis suggest that current SUSY search strategies and exclusion limits have to be reconsidered.
} %end of abstract
\maketitle
\section{Introduction}
\label{intro}
Supersymmetry (SUSY) is a popular theory of physics beyond the Standard Model (SM) because by extending the Lorentz symmetry in the only possibly way you find 
\begin{itemize}
	\item A solution to the SM hierarchy problem,
	\item A dark matter (DM) candidate,
	\item An indication of a grand unification theory (GUT)
	\item A support for String Theory
\end{itemize}
The simplest, most studied model, the constrained version of the minimal supersymmetric standard model (MSSM), is already largely excluded. Both constraints from experiments and theoretical motivations, e.g.\ the $\mu$-problem, forces us to look beyond the MSSM. To test more complicated SUSY models than the constrained MSSM or MSSM one needs to change current search strategies to make them flexible enough to be sensitive to extensions of the MSSM. The study presented here investigates the differences of gluino decays in MSSM and the $E_6$ inspired supersymmetric standard model (E$_6$SSM) \cite{Ref1}.
	\vspace{-5mm}
\section{E$_6$SSM}
\label{sec:1}
In the MSSM there is a bilinear Higgs coupling, $\mu$, which needs to be of the order 1 TeV to give an acceptable electroweak symmetry breaking. Nothing prevents this SUSY preserving coupling to be of the order of the Planck scale. To naturally get a $\mu$ of order 1 TeV one may extend the MSSM with a scalar field $S$, which couples to the Higgs fields through an interaction $\lambda S H_u H_d$ and then let $S$ get a VEV, $\langle S\rangle\equiv\frac{s}{\sqrt 2}$. This provides a $\mu$-term with $\mu=\frac{\lambda s}{\sqrt 2}$. A consequence of this is that a global $U(1)$ Peccei-Quinn symmetry is introduced and broken. The breaking of this $U(1)$ implies the existence of a massless axion, which has not been observed. There are various proposed solutions to avoid the appearance of the axion: %Various models propose solutions of how to get rid of the axion.
\begin{itemize}
	\item In the NMSSM a cubic term $S^3$ is added to break the global $U(1)$ to a descrete $Z_3$ symmetry. The breaking of this $Z_3$ could however lead to cosmological domain walls which would overclose the universe. 
	\item In the USSM the $U(1)$ is gauged and a massive $Z'$ boson appear instead but the theory is not anomaly free.
	\item In the E$_6$SSM the gauged $U(1)$ is a remnant of the breaking of a larger gauge group at the GUT scale - $E_6$. Anomalies are cancelled naturally since the particles lie in complete 27s of $E_6$.
\end{itemize}
The $E_6$ is broken down to the standard model with one extra surviving $U(1)$:
\begin{align}
	\begin{split}
	E_6&\rightarrow SO(10)\times U(1)_\psi\\
	&\rightarrow SU(5)\times U(1)_\chi\times U(1)_\psi\\
	&\rightarrow SU(3)_C\times SU(2)_W \times U(1)_Y\times U(1)_N
	\label{eq:breaking}
	\end{split}
	\nonumber
\end{align}
Each SM generation is contained in a 27 and it is the singlet, $S$, and the two Higgs doublets, $H_u$ and $H_d$, of the third 27 that are assumed to acquire VEVs. The particle content is much bigger in the E$_6$SSM than in the MSSM or USSM because of these three 27s. Contained in these 27s there are, for example, right handed neutrinos, which are neutral under the extra U(1) and can thus be heavy and provide a see-saw mechanism. Furthermore, six more, naturally light, neutralino states are introduced in addition to the six neutralinos of the USSM. These neutralinos provide a new possible source of dark matter \cite{neutralinodm} and interesting Higgs \cite{novelhiggs} and gluino phenomenology\cite{gluinopheno}.
	\vspace{-2mm}
\section{Parameter spaces}
\begin{figure}[ht]
		\centering
		\resizebox{1\columnwidth}{!}{%
		\includegraphics{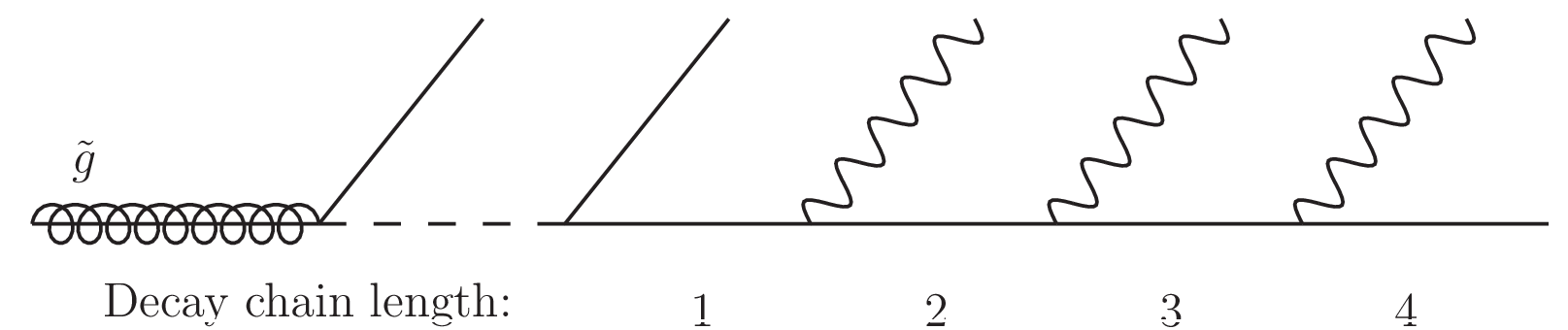}}
		\caption{The definition of the gluino decay chain length, $l$.}
		\label{fig:length}
	\end{figure}

\begin{figure}[h]
\centering
\resizebox{1\columnwidth}{!}{%
\includegraphics{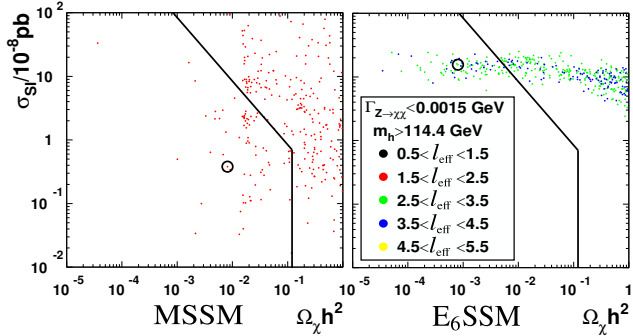}}
\caption{The scanned regions of the parameter spaces projected onto the plane spanned by the spin independent cross section, $\sigma_{SI}$, and the relic density, $\Omega h^2$. The area right of the solid line is excluded by XENON100 \cite{xenon100} and WMAP \cite{WMAP}. The colouring represents the effective gluino decay chain length $l_{\mbox{\tiny eff}}=\sum_l l\cdot P(l)$ for each point, where $P(l)$ is the probability for a chain length of $l$, as defined in Fig.\ \ref{fig:length}. The chosen benchmarks of MSSM and E$_6$SSM are encircled.}
	\label{fig:scan}
	\end{figure}

The recent XENON100\, experiment \cite{xenon100} puts a bound on the direct detection cross section for the LSP and WMAP \cite{WMAP} puts a bound on its relic density. These constraints excludes large portions of the parameter space for SUSY models. We have used CalcHEP \cite{calchep} and MicrOMEGAs \cite{micromegas} when scanning the parameter spaces of MSSM and E$_6$SSM to pick out benchmarks which satisfy these constraints on the LSP as well as constraints from collider experiments. The scanning regions are presented in Tab.\ \ref{tab:scan} and points, including benchmarks, are plotted in the plane of the LSP relic density and the direct detection cross section in Fig.\ \ref{fig:scan}. The gluino decay chain length, $l$, relevant here is defined as the number of steps in the gluino decay chain after the first squark and illustrated in Fig.\ \ref{fig:length}.
	\vspace{-7mm}
	\begin{table}
		\centering
		\begin{tabular}{|r|rr|rr|r|}
		\hline
					&	\multicolumn{2}{|c|}{MSSM}	&	\multicolumn{2}{|c|}{E$_6$SSM}&	\\	
		parameter		&	min	&	max	&	min	&	max	&	\\
		\hline
		$\tan\beta$		&	2	&	60	&	1.4	&	2	&	\\
		$|\lambda|$		&	-	&	-	&	0.3	&	0.7	&	\\
		\hline
		$s$			&	-	&	-	&	3.7	&	8	&	\multirow{4}{*}{\rotatebox{-90}{[TeV]}}\\
		$\mu$			&	-2 	&	2 	&	-	&	-	&	\\
		$A$			&	-3 	&	3 	&	-3	&	3	&	\\
		$M_A$			&	0.1 	&	2	&	1	&	5	&	\\
		\hline
		\end{tabular}
	\caption{To be able to compare the models on a common basis the gaugino masses were fixed so that a gluino mass of 800 GeV, a wino mass around 300 GeV and a bino mass around 150 GeV were provided. A common squark and slepton mass scale was fixed to $M_{S}=2$ TeV. For the E$_6$SSM a large number of Yukawa couplings were scanned over which is omitted from this table.}
	\label{tab:scan}
	\end{table}
	\vspace{-10mm}
	\section{Benchmarks}
	The benchmarks considered for MC event analysis are summarized in Tab.\ \ref{tab:bm-table}, where the neutralino spectrum is listed, and encircled in Fig.\ \ref{fig:scan}. These benchmarks do not provide a sufficient amount of dark matter and another source of dark matter is assumed. Other scenarios, where E$_6$SSM give the right amount of dark matter \cite{binodm}, have been considered in the same analysis \cite{gluinopheno}, but are not presented here. 
	\begin{table}
	\centering
	\setlength{\tabcolsep}{.5mm}
	\begin{minipage}[t!]{.6\columnwidth}
	\begin{tabular}{|c|c|c|c|c|c|c|c|c|}
		\hline
				&		MSSM	&  E$_6$SSM	&\\
		\hline
		$\tan\beta$	&		39.2	&	1.77	&\\
		$\lambda$	&		-	&	-0.462	&\\
		\hline	
		$s$		&		-	&	5418	&\multirow{4}{*}{\rotatebox{-90}{[GeV]}}\\
		$\mu$		&		1578	&	(-1770)	&\\
		$A$		&		-566.1	&	476.2	&\\
		$M_A$		&		302.5	&	2074	&\\
		$M_1$		&		150	&	150	&\\
		$M_2$		&		285	&	300	&\\
		$M_{1'}$	&		-	&	151	&\\
		$m_{\tilde g}$	&		800.2	&	800.0	&\\
		\hline
		\hline
		$P(l=1)$		&	0.188	&	$<10^{-5}$	& \\
		$P(l=2)$		&	0.812	&	0.1723		&\\
		$P(l=3)$		&	0	&	0.7986		&\\
		$P(l=4)$		&	0	&	0.02915		& \\
		$P(l=5)$		&	0	&	0		& \\
		\hline
		\hline
		$\Omega h^2$		&	0.00816	&	0.0006937	&\\
		\hline
		$\sigma_{SI}$ 
					&	$0.38\times10^{-8}$	&	$16.35\times10^{-8}$	& \rotatebox{-90}{$\!\!\!\!\!$[pb]}\\
		\hline
	\end{tabular}
		%\vspace{15.5mm}
	\end{minipage}
	\hspace{1mm}
	\begin{minipage}[T]{.34\columnwidth}
	\begin{tabular}{|c|c|c|c|c|c|c|c|c|}
		\hline
				&		MSSM	&  E$_6$SSM	&\\
		\hline
		${\tilde \chi^0_{M1}}$&		149.9	&	151.2	&\multirow{6}{*}{\rotatebox{-90}{[GeV]}}\\
		${\tilde \chi^0_{M2}}$&		302.8	&	303.7	&\\
		${\tilde \chi^0_{M3}}$&		1580	&	1766	&\\
		${\tilde \chi^0_{M4}}$&		1581	&	1771	&\\
		${\tilde \chi^\pm_{M1}}$&	302.8	&	300.9	&\\
		${\tilde \chi^\pm_{M2}}$&	1582	&	1771	&\\
		\hline
		${\tilde \chi^0_{U1}}$&		-	&	1909	&\multirow{2}{*}{\rotatebox{-90}{$\!\!\!\!\!$[GeV]}}\\
		${\tilde \chi^0_{U2}}$&		-	&	2062	&\\
		\hline	
		${\tilde \chi^0_{E1}}$&		-	&	45.2	&\multirow{8}{*}{\rotatebox{-90}{[GeV]}}\\
		${\tilde \chi^0_{E2}}$&		-	&	53.2	&\\
		${\tilde \chi^0_{E3}}$&		-	&	141.6	&\\
		${\tilde \chi^0_{E4}}$&		-	&	187.4	&\\
		${\tilde \chi^0_{E5}}$&		-	&	227.8	&\\
		${\tilde \chi^0_{E6}}$&		-	&	265.6	&\\
		${\tilde \chi^\pm_{E1}}$&	-	&	122.7	&\\
		${\tilde \chi^\pm_{E2}}$&	-	&	225.1	&\\
		\hline
		${h}$			&	119.0	&	116.3	&\multirow{2}{*}{\rotatebox{-90}{$\!\!\!\!\!$[GeV]}}\\
		${\tilde t_1}$		&	1992	&	2042	&\\
		\hline
	\end{tabular}
	\end{minipage}
		\caption{Properties of two chosen benchmarks. In the notation for neutralino and chargino states the subscript M denotes a MSSM-like state, U a USSM-like state and E an E$_6$SSM-like state. This distinction is reasonable since these sectors are very weakly coupled. The following number orders the states by mass.}
		\label{tab:bm-table}
	\end{table}
\begin{figure}[ht]
	\parbox{.2\columnwidth}{\vspace{-19mm}MSSM: (Primary chain)}
\resizebox{!}{.21\columnwidth}{%
	\includegraphics{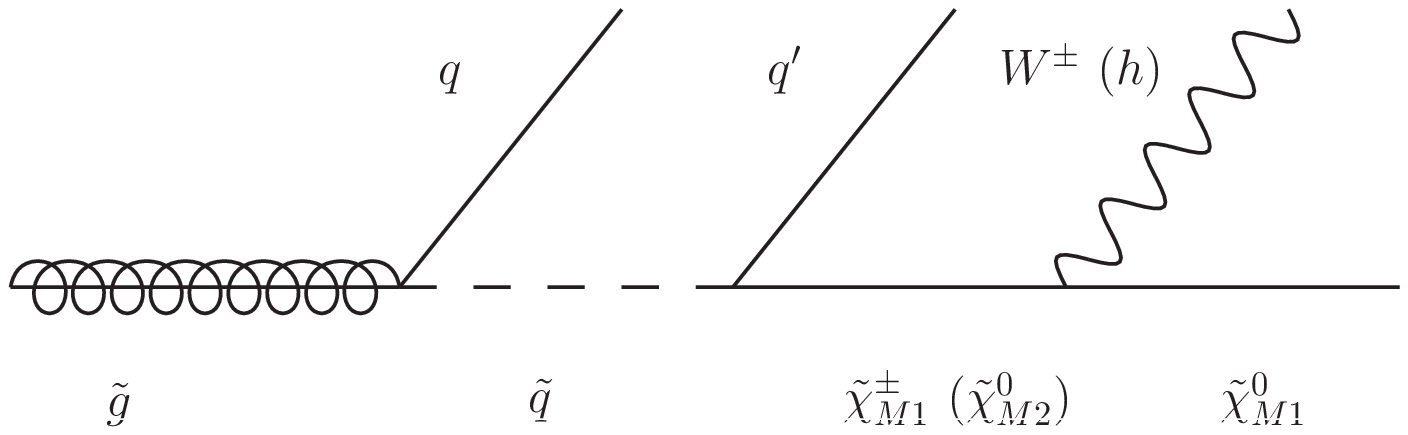}}\\
	\parbox{.2\columnwidth}{\vspace{-19mm}MSSM: (Secondary chain)}
\resizebox{!}{.21\columnwidth}{%
	\includegraphics{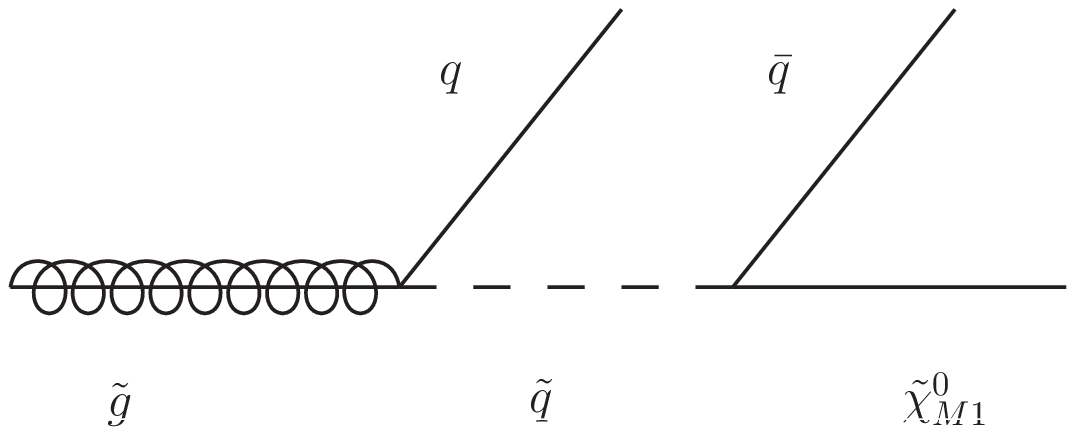}}\\
	\parbox{.2\columnwidth}{\vspace{-19mm}E$_6$SSM: (Primary chain)}
\resizebox{!}{.21\columnwidth}{%
	\includegraphics{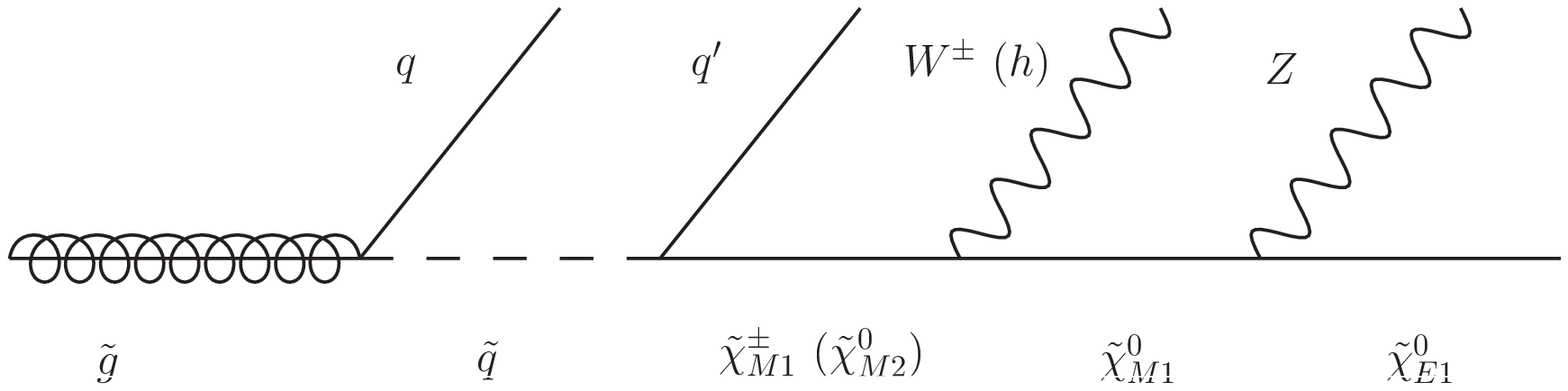}}\\
	\caption{Feynman diagrams for the leading gluino decay chains for each benchmark.}
	\label{fig:diagrams}
	%\vspace{-2mm}
\end{figure}
\section{Event analysis}
\begin{figure}[ht]
		\centering
		\resizebox{.49\columnwidth}{!}{%
		\includegraphics{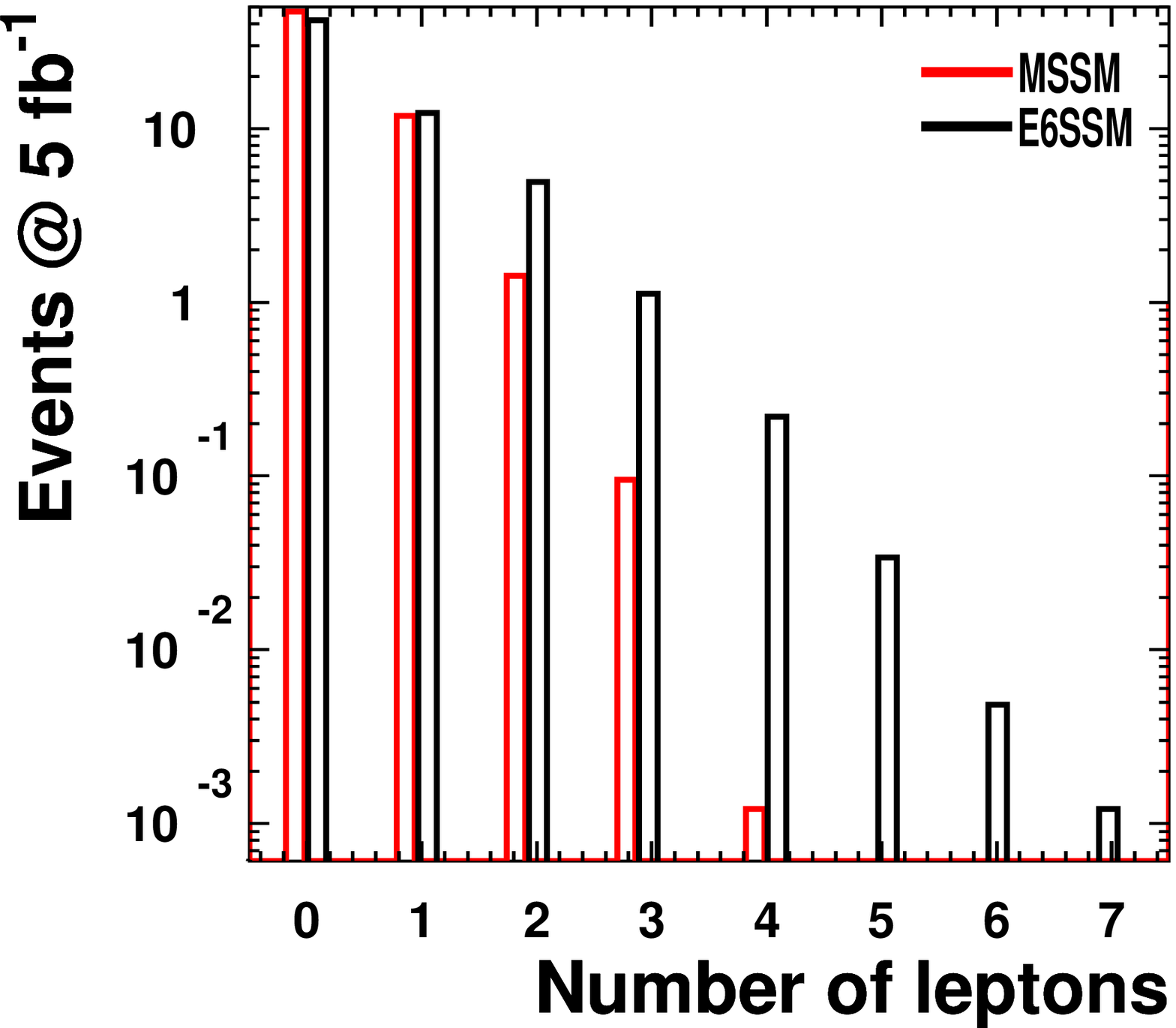}}
		\resizebox{.49\columnwidth}{!}{%
		\includegraphics{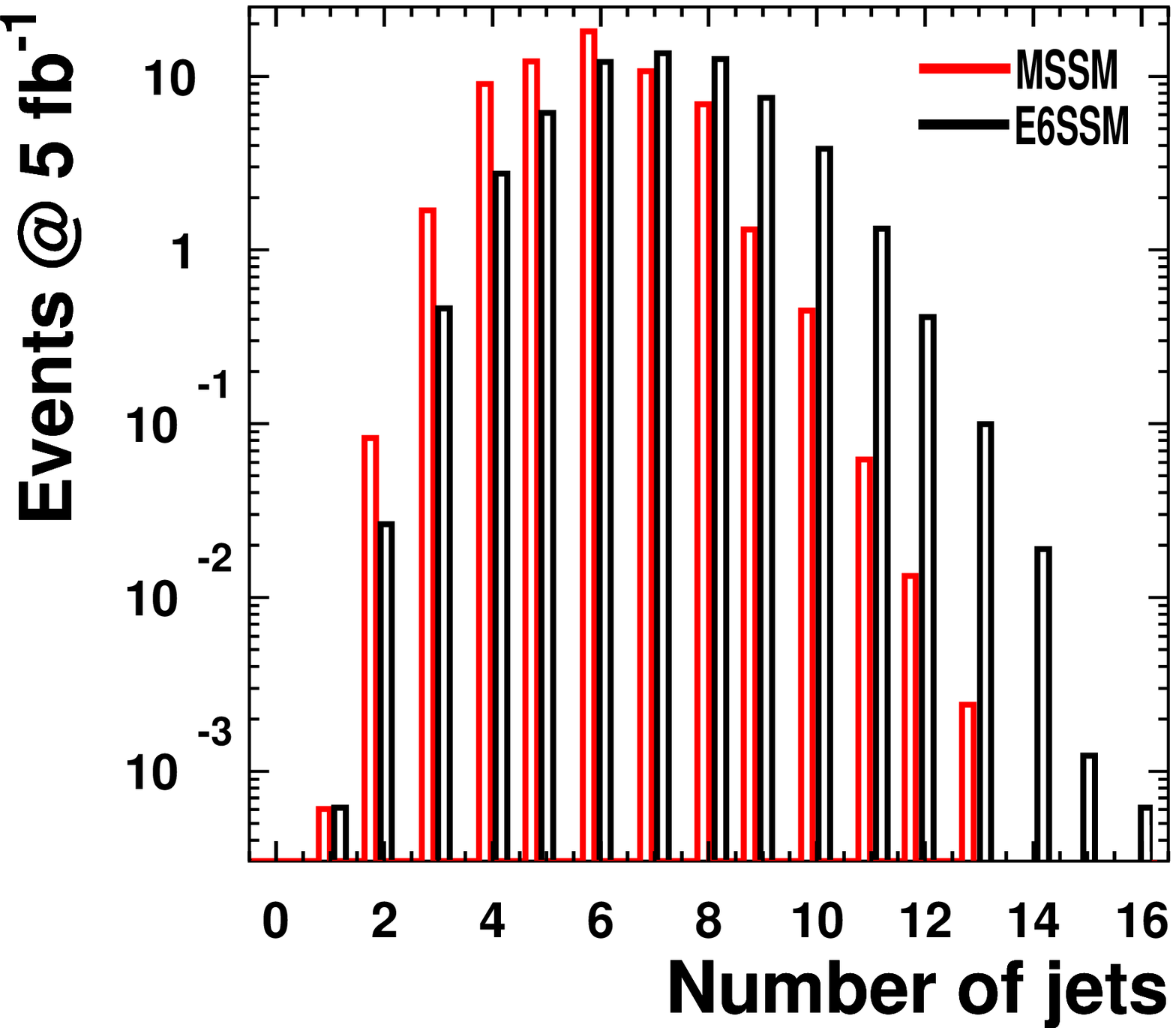}}
		\caption{Lepton multiplicity, requiring $p_T>10$ GeV and $|\eta|<2.5$, and jet multiplicity, requiring $p_T>20$ GeV and $|\eta|<4.5$.}
		\label{fig:multiplicity}
	\end{figure}

\begin{figure}[ht]
\resizebox{0.49\columnwidth}{!}{%
\includegraphics{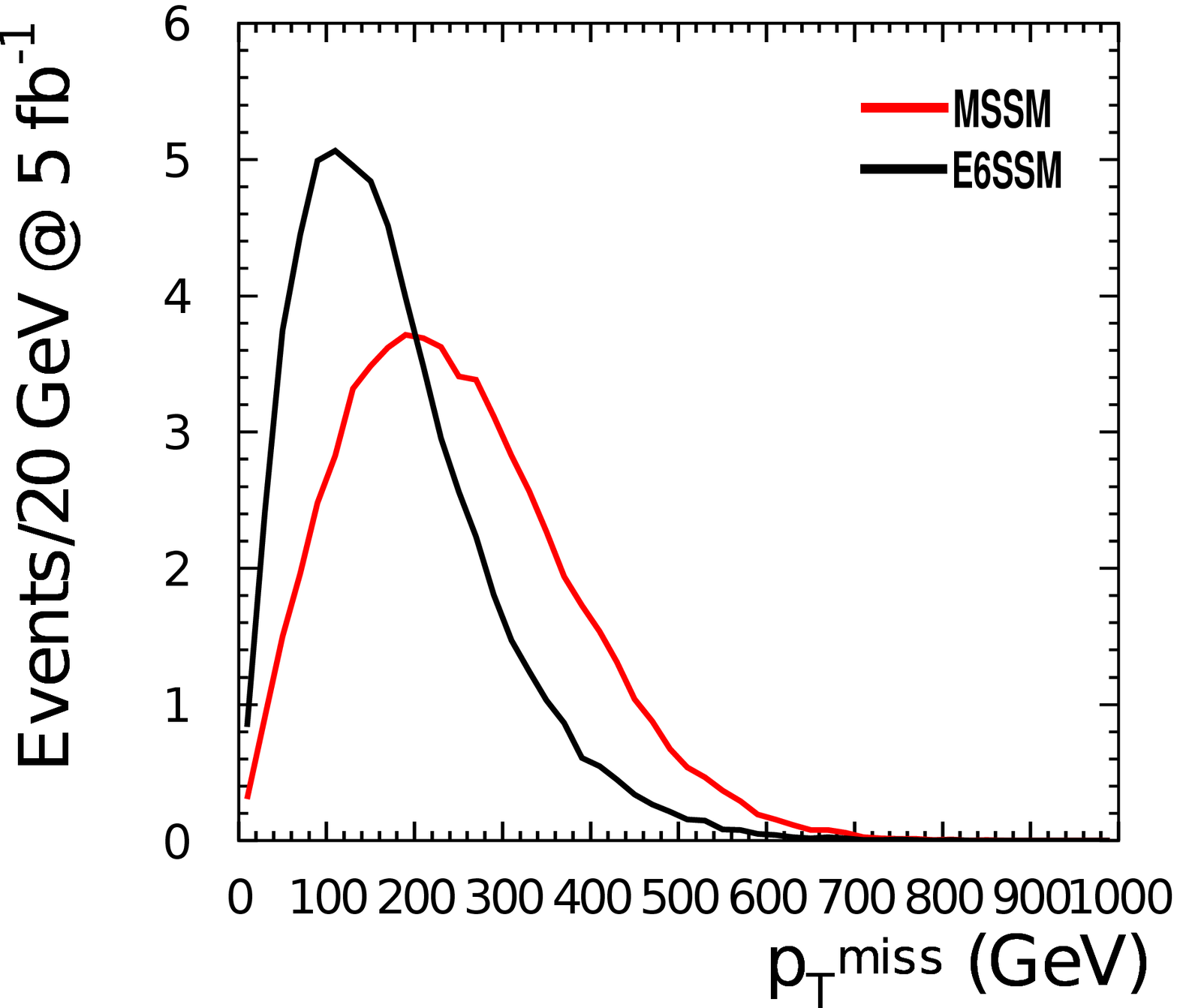}}
\resizebox{0.49\columnwidth}{!}{%
\includegraphics{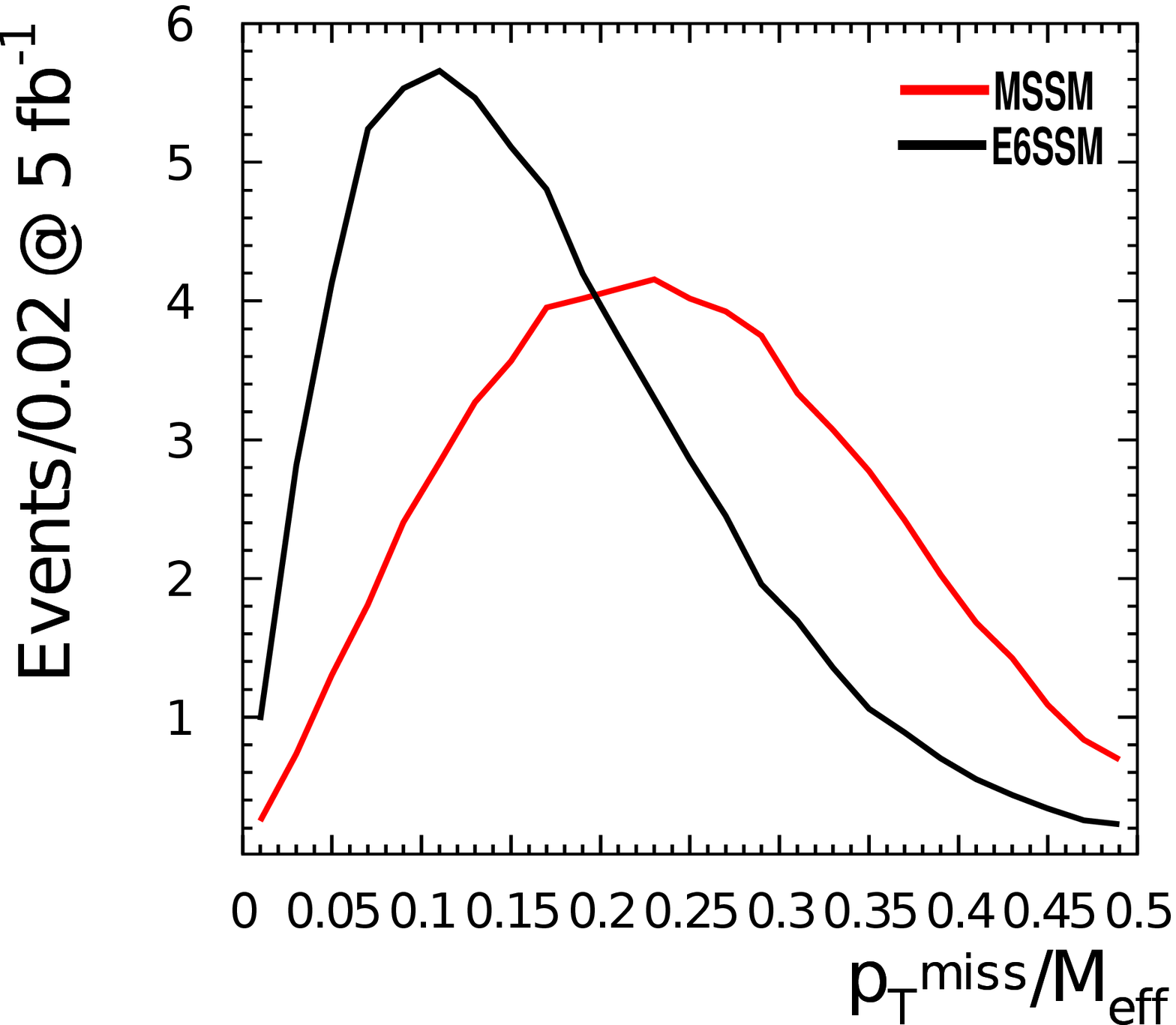}}
\caption{Missing transverse momentum and its ratio with the effective mass, $M_{\mbox{\tiny eff}}$, before selection cuts.}
		\label{fig:ptmiss}
	\end{figure}

\begin{figure}[ht]
			\centering
		\resizebox{.8\columnwidth}{!}{%
		\includegraphics{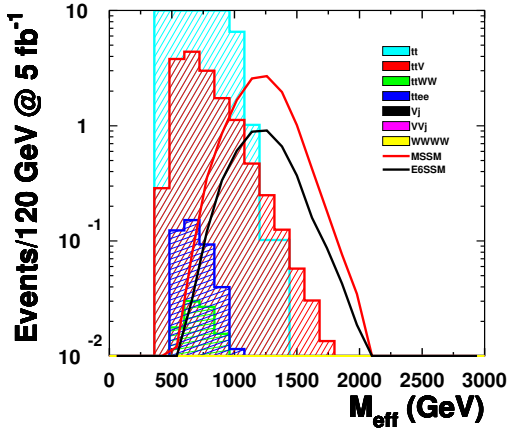}}
			\caption{The effective mass after 7 ATLAS style cuts}
			\label{fig:atlas}
		\end{figure}

\begin{figure}[ht]
			\centering
		\resizebox{.8\columnwidth}{!}{%
		\includegraphics{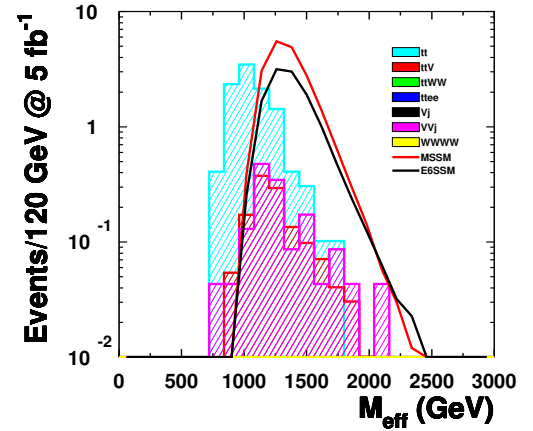}}
			\caption{The effective mass after 9 CMS style cuts}
			\label{fig:cms}
\vspace{-4mm}
		\end{figure}

Since the E$_6$SSM introduces new neutralinos, naturally lig\-ht\-er than the MSSM LSP, the gluino decay chains will be longer than the MSSM's in general. This is confirmed and illustrated by the parameter scans in Fig.\ \ref{fig:scan} and benchmarks in Tab.\ \ref{tab:bm-table}. An effect of longer decay chains is that there will be less missing momenta in collider experiments. This affects the conventional SUSY jets plus missing energy searches, e.g.\ \cite{atlas} and \cite{cms}, where the E$_6$SSM is disfavoured compared to the MSSM. This can be seen in Tab.\ \ref{tab:cuts} and Figs.\ \ref{fig:atlas} and \ref{fig:cms}. Another important feature is the increase in lepton as well as jet multiplicity, as shown in Fig.\ \ref{fig:multiplicity}. Effective variables for distinguishing models with different gluino decay chain lengths, like the MSSM and the E$_6$SSM, are $\slashed p_T$ and  $\slashed p_T/M_{\mbox{\tiny eff}}$, which are plotted in Fig.\ \ref{fig:ptmiss}, where the smaller $\slashed p_T$ and larger $M_{\mbox{\mbox{\tiny eff}}}$ of E$_6$SSM push these distributions to lower values.
%%%%%%%%%%%%%%%%%%%%%%%%%%%%%%
\vspace{-5mm}
	%%%%%%%%%%%%%%%%%%%
		\begin{table}
			\centering
\begin{tabular}{|rr|rr|rr|rr|}
\hline
\multicolumn{2}{|c|}{ATLAS CUTS} & \multicolumn{2}{|c|}{MSSM} & \multicolumn{2}{|c|}{E$_6$SSM}\\
No. & limit&Eff.	&Frac. &Eff.&Frac.\\
\hline
0 & no cut	& 0.00&	 1.00&	 0.00&	 1.00\\ 
1 & $\slashed p_T>$130 GeV	& 0.19&	 0.81&	 {0.40}&	 0.60\\ 
2 & $p_T^{jet_1}>$130 GeV	& 0.04&	 0.77&	 0.03&	 0.59\\ 
3 & $p_T^{jet_2}>$40 GeV	& 0.01&	 0.76&	 0.00&	 0.58\\ 
4 & $p_T^{jet_3}>$40 GeV	& 0.11&	 0.68&	 0.04&	 0.56\\ 
5 & $p_T^{jet_4}>$40 GeV	& 0.20&	 0.54&   0.11&	 0.50\\ 
6 & $\Delta\phi(\slashed p_T,jet)_{min}>0.4$	& 0.37&	 0.34&	{0.58}&	 0.21\\ 
7 & $\slashed p_T/M_{\mbox{\tiny eff}}>$0.25	& 0.49&	 {0.17}&	{0.68}&	 {0.07}\\ 
\hline
\end{tabular}
\vspace{1mm}
\begin{tabular}{|rr|rr|rr|}
\hline
\multicolumn{2}{|c|}{CMS CUTS}  & \multicolumn{2}{|c|}{MSSM} & \multicolumn{2}{|c|}{E$_6$SSM}\\
No. & limit	&Eff.	&Frac.	&Eff.	&Frac.\\
\hline
0 & no cut		& 0.00&	 1.00&	 0.00&	 1.00\\ 
1 & $\slashed H_T>200$ GeV	& 0.34&	 0.66&	 {0.47}&	 0.53\\ 
2 & $p_T^{jet_1}>$50 GeV	& 0.00&	 0.66&	 0.00&	 0.53\\ 
3 & $p_T^{jet_2}>$50 GeV	& 0.02&	 0.64&	 0.01&	 0.53\\ 
4 & $p_T^{jet_3}>$50 GeV	& 0.13&	 0.56&	 0.06&	 0.50\\ 
5 & $\Delta\phi(\slashed p_T,jet_1)>0.5$	& 0.02&	 0.55& 0.03&	 0.48\\ 
6 & $\Delta\phi(\slashed p_T,jet_2)>0.5$	& 0.08&	 0.50&	 {0.12}&	 0.42\\ 
7 & $\Delta\phi(\slashed p_T,jet_3)>0.3$	& 0.07&	 0.47&	 {0.10}&	 0.38\\ 
8 & $\Delta R (jet,lep)_{min}<0.3$	&0.24&	 0.36&	 0.36&	 0.25\\ 
9 & $H_T>800$ GeV	&0.49&	 {0.18}&	 0.38&	 {0.15}\\ 
\hline
\end{tabular}
			\caption{Efficiency of ATLAS and CMS cuts and the fraction events left after successive applications of cuts.}
			\label{tab:cuts}
\vspace{-5mm}
		\end{table}
\section{Conclusions}
Careful analysis has to be made to distinguish SUSY models. The models studied here are very different but conventional cuts and the use of effective mass makes them blend into each other. The E$_6$SSM has large visible and small missing $p_T$. The effect of these features cancels in $M_{\mbox{\tiny eff}}$, while it is enhanced in $\slashed p_T/\sum_{\mbox{\tiny visible}}|p_T^{\mbox{\tiny visible}}|$. Hard cuts on $\slashed p_T$ and $\slashed p_T/M_{\mbox{\tiny eff}}$ or equivalents, like those used by ATLAS and CMS in Tab.\ \ref{tab:cuts}, are severe for models with long decay chains. Large jet and lepton multiplicity searches should be the way forward for models like the E$_6$SSM.
	\vspace{3mm}\\
\noindent	
	{\bf Acknowledgements}\\
	We would like to thank Alexander Pukhov for necessary improvements of CalcHEP and MicroMEGAs. PS thanks the NExT institute and SEPnet for support. AB thanks the NExT Institute and Royal Society for partial financial support.
\vspace{-2mm}
%\begin{figure}
% Use the relevant command for your figure-insertion program
% to insert the figure file.
% For example, with the option graphics use
%\resizebox{0.75\columnwidth}{!}{%
%  \includegraphics{fig1.eps} }
%\caption{Please write your figure caption here.}
%\label{fig:1}       % Give a unique label
%\end{figure}
%
% For tables use
%\begin{table}
%\caption{Please write your table caption here.}
%\label{tab:1}       % Give a unique label
%% For LaTeX tables use
%\begin{tabular}{lll}
%\hline\noalign{\smallskip}
%first & second & third  \\
%\noalign{\smallskip}\hline\noalign{\smallskip}
%number & number & number \\
%number & number & number \\
%\noalign{\smallskip}\hline
%\end{tabular}
%\end{table}
%

\end{document}